\documentclass[aps,onecolumn,showpacs,amsmath,amssymb]{revtex4}
\usepackage[dvips]{graphics}
\begin{document}
\large
\title{$q$-breathers in Discrete Nonlinear Schr\"{o}dinger lattices}
\author{K.~G.~Mishagin$^1$, S.~Flach$^2$, O.~I.~Kanakov$^1$ and M.~V.~Ivanchenko$^{1,3}$}
\affiliation{ $^1$ Department of Radiophysics, Nizhny Novgorod
University, Gagarin Avenue, 23, 603950 Nizhny Novgorod, Russia
\\
$^2$ Max Planck Institute for the Physics of Complex Systems,
N\"othnitzer Str. 38, D-01187 Dresden, Germany
\\
$^3$ Department of Applied Mathematics, University of Leeds, Leeds
LS2 6JT, United Kingdom}



\date{\today}

\begin{abstract}
$q$-breathers are exact time-periodic solutions of extended
nonlinear systems continued from the normal modes of the
corresponding linearized system. They are localized in the space
of normal modes. The existence of these solutions in a weakly
anharmonic atomic chain explained essential features of the
Fermi-Pasta-Ulam (FPU) paradox.
We study $q$-breathers in one- two- and
three-dimensional discrete nonlinear Sch\"{o}dinger (DNLS)
lattices --- theoretical playgrounds for light propagation in
nonlinear optical waveguide networks, and the dynamics of cold
atoms in optical lattices.
We prove the existence of these solutions for weak nonlinearity.
We find that the localization of
$q$-breathers is controlled by a single parameter which depends on
the norm density, nonlinearity strength and seed wave vector. At a
critical value of that parameter $q$-breathers delocalize via
resonances, signaling a breakdown of the normal mode picture and
a transition into strong mode-mode interaction regime. In
particular this breakdown takes place at one of the edges of the
normal mode spectrum, and in a singular way also in the center of
that spectrum. A stability analysis of $q$-breathers supplements
these findings. For three-dimensional lattices, we find $q$-breather
vortices, which violate time reversal symmetry and generate a vortex ring
flow of energy in normal mode space.
\end{abstract}


\maketitle

\section{Introduction}

\subsection{Background}

In 1955, Fermi, Pasta and Ulam (FPU) published their seminal report
on the absence of thermalization in arrays of particles connected
by weakly nonlinear springs \cite{fpu}. In particular they observed
that energy, initially seeded in a low-frequency normal mode of the linear problem
with a frequency $\omega_q$ and corresponding normal mode number $q$, stayed
almost completely locked in a few neighbor modes in frequency space, instead
of being distributed quickly among all modes of the system.
The latter expectation was due to the fact, that nonlinearity does induce
a long-range network of interactions among the normal modes.
>From the present
perspective, the FPU problem appears to consist of the following parts:
(i) for certain parameter ranges (energy, system size, nonlinearity strength,
seed mode number), excitations appear to stay {\sl exponentially localized} in $q$-space of the
normal modes
for {\sl long but finite} times \cite{lgas72};
(ii) this {\sl intermediate} localized state is reached on a fast time scale $\tau_1$,
and {\sl equipartition is reached} on a second time scale $\tau_2$, which can be many orders
of magnitude larger than $\tau_1$ \cite{GIORGILLI SHOWERS};
(iii) tuning the control parameters may lead to
a drastical shortening of $\tau_2$, until both time scales merge, and the metastable
regime of localization is replaced by a fast relaxation towards thermal equilibrium
\cite{GIORGILLI SHOWERS},
which is related to the nonlinear resonance overlap estimate by Izrailev and Chirikov
\cite{fmibvc66}.
FPU happened to compute cases with values of $\tau_2$ which were inaccessible within their
computation time. Among many other intriguing details of the evolution of the
intermediate localized state, we mention the possibility of having an almost regular
dynamics of the few strongly excited modes, leading to the familiar phenomenon of beating.
That beating manifests as a recurrence of almost all energy into the originally excited
mode (similar to the beating among two weakly coupled harmonic oscillators),
and was explained by Zabusky and Kruskal with the help of real space solitons
of the Kordeweg-de-Vries (KdV) equation,
for a particular case of long wavelength seed modes, and periodic boundary conditions
\cite{njzmdk65}.
The regular dynamics may be replaced by a weak chaotic one upon crossing yet another
threshold in control parameters \cite{jdlalmal95}. Remarkably, that weak chaos is still confined
to the
core of the localized excitation, leaving the exponential localization almosts unchanged, and
perhaps
influencing most strongly the parameter dependence of $\tau_2$. The interested reader may also
consult Ref.\cite{fpumore}.

\subsection{Motivation}

Single-site excitations in translationally invariant lattices of interacting anharmonic
oscillators are known to show
a similar behaviour of trapping the excitation on a few lattice sites around the originally
excited one
\cite{dbnumerics}. As conjectured almost 40 years ago \cite{ovchinnikov}, exact time-periodic
and
spatially localized orbits - {\sl discrete breathers} (also {\sl intrinsic localized modes,
discrete solitons})
persist in such lattices, and the dynamics on such an orbit, as well as the dynamics of the
nearby phase space
flow, account for many observations, and have found their applications in many different areas
of physics
\cite{db}. Recently, it was shown, that the regime of localization in normal mode space for the
FPU problem
can be equally explained by obtaining {\sl $q$-breathers} (QB) - time-periodic and
modal-space-localized orbits -
which persist in the FPU model for nonzero nonlinearity \cite{QB}.
The originally computed FPU trajectory
stays close to such $q$-breather solutions for long times, and therefore many features
of its short- and medium-time dynamics have been shown to be captured by the $q$-breather
solution and small phase-space fluctuations around it.
Delocalization thresholds for $q$-breathers are related to resonances, and corresponding
overlap criteria \cite{QB}.
The method of constructing $q$-breather solutions was generalized to two- and three-dimensional
FPU lattices \cite{FPU2d3d}. Most importantly, a scaling theory was developed, which allowed
to construct $q$-breathers for arbitrarily large system sizes, and to obtain analytical
estimates on the degree of localization of these solutions for macroscopic system size
\cite{Scaling}. Expectations were formulated, that the existence of $q$-breathers
should be generic to many nonlinear spatially extended systems.

\subsection{Aim}

The spectrum of normal mode frequencies of the linear part of the FPU model is acoustic,
contains zero, reflecting the fact, that the model 
conserves total mechanical momentum. In addition it is bounded by
a lattice-induced cutoff, the analogue of the Debye cutoff in solid state physics.
We may think of two pathways of extending the above discussed results to other system
classes. First, we could consider a spatially continuous system, instead of a lattice.
However, the initial FPU studies showed confinement to a few long wavelength
modes, and the results of Zabusky and Kruskal \cite{njzmdk65},
confirm, that indeed in the spatially continuous system (KdV), trajectories similar to
the FPU trajectory persist.

As discussed above, the persistence, localization, and delocalization of $q$-breathers
is due to a proper avoiding of, or harvesting on, resonances between normal modes.
It therefore appears to be of substantial interest to extend the previous results
to systems with a qualitatively different normal mode spectrum.
Such a spectrum is one which does not contain zero, and is called optical.
It may correspond to the excitation of certain degrees of freedom in a complex
system, or to an FPU type model which is deposited on some substrate, or otherwise
exposed to some external fields.
Together with the acoustic band, these two band structures, when combined, describe
almost any type of normal mode spectrum in a linear system with spatially periodic
modulated characteristics.

Adding weak nonlinearities to such a system, taking local action
angle representations, and performing various types of multiple
scale analysis, such models are mapped on discrete nonlinear
Schr\"odinger models (DNLS) \cite{DNLS derivations}. These models
enjoy a gauge invariance, and a conservation of the sum of the
local actions (norms), i.e. some global norm. A particular version
of these equations is known as the discrete Gross-Pitaevsky equation, and
is derived on mean field grounds for Bose-Einstein condensates of
ultracold bosonic atoms in optical lattices \cite{grosspitaevsky,healinglength}. The
norm is simply the conserved number of atoms in this case, and the
nonlinearity derives from the atom-atom interaction. The
propagation of light in (spatially modulated) optical media is
another research area where DNLS models are used
\cite{nonlinearoptics}. In that case the norm conservation derives
from the conservation of electromagnetic wave energy along the
propagation distance (in the assumed absence of dissipative
mechanisms). But DNLS models served equally well in many other
areas of physics, where weak interactions start to play a role,
e.g. in the theory of polaron formation due to electron-phonon
interaction in solids.

Releasing the norm conservation (e.g. by allowing the number of atoms in a condensate
to fluctuate) will reduce the symmetries of the corresponding model, but most
importantly, it will lead to possible new resonances of higher harmonics.
We will discuss these limitations, and possible effects of releasing these limitations
on $q$-breathers, in the discussion section.

A particular consequence of norm conservation is a corresponding symmetry of the
normal mode spectrum, which relates both (upper and lower) band edges, and
makes the band center a symmetry point as well.

The paper is structured as follows.
Section \ref{Model} is introducing the model and the main equations of motion.
Section \ref{existence} gives an existence proof for $q$-breathers.
In section \ref{perturbation} we derive analytical expressions for the
QB profiles using perturbation theory. These are compared with the numerical
results for $d=1$ in section \ref{results1d}, which also contains a stability analysis
of the obtained periodic orbits.
Section \ref{pbc} gives results on periodic boundary conditions (as opposed to
fixed boundaries). We generalize to $d=2,3$ in section \ref{results2d3d}, and
discuss all results in section \ref{discussion}.

\section{Model}
\label{Model}
We consider a discrete nonlinear Schr\"odinger (DNLS) equation on a $d$-dimensional hypercubic  lattice
of linear size
$N$:
\begin{equation}\label{eq1}
i\dot{\psi_{\boldsymbol{n}}}=\sum\limits_{\boldsymbol{m}\in
D(\boldsymbol{n})}\psi_{\boldsymbol{m}}+\mu
|\psi_{\boldsymbol{n}}|^2 \psi_{\boldsymbol{n}}.
\end{equation}
Here $\psi$ is a complex scalar which may describe e.g. the
probability amplitude of an atomic cloud on an optical lattice
site \cite{healinglength}, or relates to the amplitudes of a propagating
electromagnetic wave in an optical waveguide \cite{nonlinearoptics}. The
lattice vectors $\boldsymbol{n, m}$ have integer components, and
$D(\boldsymbol{n})$ is the set of nearest neighbors for the
lattice site $\boldsymbol{n}$. If not noted otherwise, we consider
fixed boundary conditions: $\psi_{\boldsymbol{n}}=0$ if $n_l=0$ or
$n_l=N+1$ for any of the components of $\boldsymbol{n}$. Equation
(\ref{eq1}) is derived from the Hamiltonian
\begin{equation}\label{eq2}
  H=\sum\limits_{\boldsymbol{n}}
  \left(\sum\limits_{\boldsymbol{m}\in D(\boldsymbol{n})}
  \psi_{\boldsymbol{m}}\psi_{\boldsymbol{n}}^{*}+\frac{\mu}{2}|\psi_n|^4\right)
\end{equation}
using the equations of motion $i\dot{\psi_n}=\partial
H/\partial\psi_{n}^{*}$. In addition to energy, equation
(\ref{eq1}) conserves the norm $B=\sum\limits_{\boldsymbol{n}}
|\psi_{\boldsymbol{n}}|^2$. Note that the change of the
nonlinearity parameter $\mu$ in (\ref{eq1}) is strictly equivalent
to changing the norm $B$. Here we will keep the norm $B$
(alternatively the norm density) fixed,  and vary $\mu$.

We perform a canonical transformation to the reciprocal space of normal
modes ($q$-space of size $N^d$) with new variables
$Q_{\boldsymbol{q}}(t)\equiv Q_{q_1\ldots q_d}(t)$
\begin{equation}\label{eq3}
\psi_{\boldsymbol{n}}(t)=\left(\frac{2}{N+1}\right)^{d/2}\sum\limits_{q_1,\ldots,
q_d=1}^N Q_{\boldsymbol{q}}(t)\prod\limits_{i=1}^d
\sin{\left(\frac{\pi q_i n_i}{N+1}\right)}\;.
\end{equation}
Together with (\ref{eq1}) we obtain the following equations of
motion in normal mode space:
\begin{equation}\label{eq4}
  i\dot{Q_{\boldsymbol{q}}}=-\omega_{\boldsymbol{q}}Q_{\boldsymbol{q}} +
\frac{2^{d-2}\mu}{(N+1)^d}
  \sum\limits_{\boldsymbol{p,r,s}} C_{\boldsymbol{q,p,r,s}}
  Q_{\boldsymbol{p}}Q_{\boldsymbol{r}}^*Q_{\boldsymbol{s}},
\end{equation}
where $\omega_{\boldsymbol{q}}=-2 \sum\limits_{i=1}^d
\cos{\frac{\pi q_i}{N+1}}$ are the normal mode frequencies for the
linearized system (\ref{eq1}) with $\mu=0$. Nonlinearity
introduces a network of interactions among the normal mode
oscillators with the following coupling coefficients:
\begin{equation}\label{eq5}
 \begin{array}{c}
  C_{\boldsymbol{q,p,r,s}}=\prod\limits_{i=1}^{d}
  C_{q_i,p_i,r_i,s_i}, \\
  C_{q_i,p_i,r_i,s_i}=\sum\limits_{k,l,m=0}^{1}
  (-1)^{k+l+m} (\delta_{(-1)^k p_i+(-1)^l r_i+(-1)^m s_i,q_i}
  + \\ +\delta_{(-1)^k p_i+(-1)^l r_i+(-1)^m s_i,q_i \pm (2N+2)}).
 \end{array}
\end{equation}

\section{Proof of existence of $t$-reversible $q$-breather solutions for weak nonlinearity}
\label{existence}

We look for exact time-periodic solutions, which are stationary
solutions of the DNLS equation (\ref{eq1}), and which are
localized in normal mode space:
$\psi_{\boldsymbol{n}}(t)=\phi_{\boldsymbol{n}} exp(i\Omega t)$
with frequency $\Omega$ and time-independent amplitudes
$\phi_{\boldsymbol{n}}$. In the space of normal modes these
stationary solutions have the form
$Q_{\boldsymbol{q}}(t)=A_{\boldsymbol{q}}exp(i\Omega t)$, where
the amplitudes of modes $A_{\boldsymbol{q}}$ are time-independent
and related to the real-space amplitudes by the transformation
(\ref{eq3}). At a given norm $B$ they satisfy a system of algebraic equations:
\begin{equation}\label{eq6}
\left\{
\begin{array}{lc}
-\Omega A_{\boldsymbol{q}}+\omega_{\boldsymbol{q}}
A_{\boldsymbol{q}}-\frac{2^{d-2}\mu}{(N+1)^d}
\sum\limits_{\boldsymbol{p},\boldsymbol{r},\boldsymbol{s}}
C_{\boldsymbol{q},\boldsymbol{p},\boldsymbol{r},\boldsymbol{s}}
A_{\boldsymbol{p}}A^*_{\boldsymbol{r}}A_{\boldsymbol{s}} = 0,
& q_{1,..,d}=\overline{1,N},\\
\sum\limits_{\boldsymbol{q}} |A_{\boldsymbol{q}}|^2 - B = 0, &
\end{array}
\right.
\end{equation}
We are focusing here (and throughout almost all of
the paper) on $t$-reversible periodic orbits. Therefore we may
consider all $A_{\boldsymbol{q}}$ to be real numbers. In this case system (\ref{eq6}) 
contains $N+1$ equations for $N+1$ variables. This system can be condensed into an equation 
for a vector function:
\begin{equation}\label{eq7}
\boldsymbol{F}(\boldsymbol{X};\mu,B)=0
\end{equation}
with $\boldsymbol{X}=\{\ldots,A_{\boldsymbol{q}},\ldots,\Omega\}$.
The components of $\boldsymbol{F}$ are the left hand sides of
(\ref{eq6}), while $\mu$, $B$ are parameters.

For $\mu=0$ the normal modes in (\ref{eq4}) are decoupled and each
oscillator conserves its norm in time:
$B_{\boldsymbol{q}}(t)=A_{\boldsymbol{q}}^2$. Let us consider the
excitation of only one of the oscillators with the seed mode
number $\boldsymbol{q}_0$:
$B_{\boldsymbol{q}}=B_{\boldsymbol{q}_0}\delta_{\boldsymbol{q},\boldsymbol{q}_0}$.
The excited normal mode is a time-periodic solution of
(\ref{eq4}), and is localized in $\boldsymbol{q}$-space. According
to the Implicit Function Theorem \cite{ImpFunc}, the corresponding
solution of (\ref{eq6}) can be continued into the nonlinear case
($\mu\neq 0$), if the Jacoby matrix of the linear solution
$(\partial\boldsymbol{F}/\partial\boldsymbol{X})_{\mu=0,B}$ is
invertible, i.e.
$||\partial\boldsymbol{F}/\partial\boldsymbol{X}||_{\mu=0,B}\neq
0$. The Jacobian
\begin{equation}\label{eqJ}
 \left\|{\frac{\partial\mathbf{F}}{\partial\mathbf{X}}}\right\|_{\mu=0}=
 (-1)^{N^d+1}
2A_{\boldsymbol{q}_0}^2\prod\limits_{\boldsymbol{q}\neq\boldsymbol{q}_0}
 (\omega_{\boldsymbol{q}}-\omega_{\boldsymbol{q}_0}).
\end{equation}
Invertibility of the Jacoby matrix requires a non-degenerated
spectrum of normal mode frequencies. Therefore, the continuation
of a linear mode with the seed mode number $\boldsymbol{q}_0$ will
be possible if
$\omega_{\boldsymbol{q}}\neq\omega_{\boldsymbol{q}_0}$, for all
$\boldsymbol{q}\neq\boldsymbol{q}_0$. That condition holds for the
case $d=1$, and thus $q$-breathers exist at least for suitably
small values of $\mu$ there. For higher dimensions, degeneracies
of normal mode frequencies appear. These degeneracies are not an
obstacle for numerical continuation of $q$-breathers, but a formal
persistence proof has to deal with them accordingly. Analogous
results for two- and three-dimensional $\beta$-FPU lattices have
been obtained in Ref. \cite{FPU2d3d}.

\section{Perturbation theory for $q$-breather profiles}
\label{perturbation}

To analyze the localization properties of $q$-breathers in
$q$-space we use a perturbation theory approach similar to
\cite{QB},\cite{FPU2d3d}. We consider the general case of a
$d$-dimensional DNLS lattice (\ref{eq1}). Taking the solution for
a linear normal mode with number
$\boldsymbol{q}_0=(q_{1,0},\ldots,q_{d,0})$ as a zero-order
approximation, an asymptotic expansion of the solution to the
first $N$ equations of (\ref{eq6}) in powers of the small
parameter $\sigma=\mu 2^{(d-2)}/(N+1)^d$ is implemented. Note,
that in the same way as it was done in \cite{QB},\cite{FPU2d3d},
we fix the amplitude of the seed mode $A_{\boldsymbol{q}_0}$.
Later on we will apply the norm conservation law (the last
equation of (\ref{eq6})) to express the norm
$B_{\boldsymbol{q}_0}=|A_{\boldsymbol{q}_0}|^2$ via the total norm
$B$ for the case of $d=1$. Analytical estimations presented below
describe amplitudes of modes located along the directions of the
lattice axes starting from the mode $\boldsymbol{q}_0$. Mode
amplitudes of $q$-breathers have the slowest decay along these
directions. Studying $q$-breather localization along a chosen
dimension $i$, for the sake of compactness we will use a scalar
mode number to denote the $i$-th component of $\boldsymbol{q}$,
assuming all other components be the same as in
$\boldsymbol{q}_0$: $q_{j\neq i}=q_{j,0}$.

\subsection{Close to the band edges}

Let us start with $q$-breathers localized in the low-frequency
mode domain ($q_{i,0}<<N$, $i=\overline{1,d}$). According to the
selection rules, if the seed mode number $q_{i,0}$ is even (odd),
then only even (odd) modes are excited along $i$-th dimension. The
$n$-th order of the asymptotic expansion is the leading one for
the mode $q_{i,n}=(2n+1)q_{i,0}$. When $q_i$ reaches the upper
band edge in the $i$-th direction, a reflection at the edge in
$q$-space takes place if $N+1$ is not divisible by $q_{i,0}$ (cf.
Fig. \ref{fig1}d). If $N+1$ is divisible by $q_{i,0}$ then only
modes $q_{i,n}=(2n+1)q_{i,0}<N$ are excited. In the analytical
estimates we assume a large enough lattice size. Then the effect
of band edge reflections is appearing in higher orders of the
perturbation, which will not be considered. In this case the approximate
solution is
\begin{equation}
A_{(2n+1)q_{i,0}}=(-\mbox{sign}(\mu))^n\left(\sqrt{\lambda_d^{(i)}}\right)^{n}
A_{\boldsymbol{q}_0}\;,\; i=\overline{1,d}\;,\;
\Omega=\omega_{\boldsymbol{q}_0}-\sigma A_{\boldsymbol{q}_0}^2 +
O(\sigma^2)\;. \label{solutionloweredge}
\end{equation}
The corresponding exponential decay of mode norms is
\begin{equation}\label{eq8}
B_{(2n+1)q_{i,0}}=(\lambda_d^{(i)})^n B_{\boldsymbol{q}_0}\;,\;
  \sqrt{(\lambda_d^{(i)})}=\frac{|\mu| A_{\boldsymbol{q}_0}^2 (N+1)^{2-d}}
  {2^{5-d} \pi^2{(q_{i,0})}^2} = \frac{|\mu| b_{\boldsymbol{k}_0}}
  {2^{5-d} {(k_{i,0})}^2}\;,\;
  i=\overline{1,d}\;.
\end{equation}
$\lambda_d^{(i)}$ ($0<\lambda_d^{(i)}<1$) characterize the
exponential decay along the $i$-th dimension. Here
$k_{i,0}=\frac{\pi q_{i,0}(2n+1)}{N+1}<\pi$ is the $i$-th
component of the seed wave vector $\boldsymbol{k}_0$, and
$b_{\boldsymbol{k}_0}=B_{\boldsymbol{k}_0}/(N+1)^d$ is the norm
density of the seed mode. When using intensive quantities -- wave
number $k_{i,0}$ and norm density $b_{\boldsymbol{k}_0}$ -- only,
$\lambda$ does not depend on the system size.

Equations (\ref{eq4}) are invariant under the symmetry operation
$\mu \rightarrow -\mu$, $t \rightarrow -t$, $q_{i,0} \rightarrow
N+1 - q_{i,0}$ for all $i=\overline{1,d}$, which changes the sign
of nonlinearity, and maps modes from one band edge to the other.
The replacement $q_{i,0} \rightarrow N+1 - q_{i,0}$ for all
$i=\overline{1,d}$ in (\ref{eq6}) is equivalent to substitutions
$\mu \rightarrow -\mu$ and $\Omega \rightarrow -\Omega$. Using
this symmetry, we can easily apply the above results to the case
of $q$-breathers localized near the upper band edge, by counting
mode indices from the upper edge: $\widetilde{q}_i=N+1-q_i$. We
neglect reflections from the lower band edge, such that only modes
with numbers $\widetilde{q}_{i,n}=(2n+1)\widetilde{q}_{i,0}$ ($n$
is integer, $\widetilde{q}_{i,0}<<N$) are assumed to be excited.
It follows
\begin{equation}\label{solutionupperedge}
A_{N+1-\widetilde{q}_{i,n}}=(\mbox{sign}(\mu))^n\left(\sqrt{\widetilde{\lambda}_d^{(i)}}\right)^n
A_{\boldsymbol{q}_0}
\end{equation}
where $n$ is an integer and
\begin{equation}\label{eq8_2}
 \begin{array}{cc}
  \sqrt{(\widetilde{\lambda}_d^{(i)})}=\frac{|\mu| A_{\boldsymbol{q}_0}^2 (N+1)^{2-d}}
  {2^{5-d} \pi^2(N+1-q_{i,0})^2} = \frac{|\mu| b_{\boldsymbol{k}_0}}
  {2^{5-d} (\pi-k_{i,0})^2}\;,\; &
  i=\overline{1,d}\;.
 \end{array}
\end{equation}
The analytical expression for the frequency of the $q$-breather
solution turns out to be the same as for the case of small seed mode numbers.

\subsection{Close to the band center}

We implement the perturbation theory approach for seed modes close
to the band center.
Let us first consider the case of odd $N$. We introduce the new index
$p^i$, which is the number of a mode counted from the middle of the
spectrum along the $i$-th dimension: $p_i=q_i-(N+1)/2$.
We choose the seed mode with
$|p_{i,0}|<<(N+1)/2$.
In the $n$-th order of perturbation theory
the newly excited mode along the $i$-th dimension has
the number $p_{i,n}=(-1)^{n}(2n+1)p_{i,0}$, and
\begin{equation}\label{eq13}
 \begin{array}{l}
  A_{p_{i,2n}}=(-{\lambda_d^{(i)}})^n A_{p_{i,0}}\;,\;
  A_{p_{i,2n+1}}=\mbox{sign}(p_{i,0})(-1)^{n+1}(\lambda_d^{(i)})^{n+1/2} A_{p_{i,0}}\;,
 \end{array}
\end{equation}
where $i=\overline{1,d}$, $\lambda_d^{(i)}>0$, $n=1,2,3,\ldots$ , and
\begin{equation}\label{eq14}
\sqrt{\lambda_d^{(i)}} =
\frac{|\mu|A_{\boldsymbol{q}_0}^2}{2^{4-d}\pi
|q_{i,0}-(N+1)/2|}=\frac{|\mu| b_{\boldsymbol{k}_0}}{2^{4-d}
|k_{i,0}-\pi/2|}.
\end{equation}
When using intensive quantities, $\lambda$ again does not depend
on the system size.

For the case of even $N$ we use $p_i=q_i-N/2$ and assume that the
seed mode index $p_{i,0}>0$, $i=\overline{1,d}$. The set of
consecutively perturbed modes becomes more complicated:
\begin{equation}\label{eq15}
p_{i,0} \rightarrow (-3p_{i,0}+2) \rightarrow (5p_{i,0}-2)
\rightarrow (-7p_{i,0}+4) \rightarrow (9p_{i,0}-4) \rightarrow
(-11p_{i,0}+6) \rightarrow \ldots \;.
\end{equation}
For an even number of perturbation steps the new excited mode along the $i$-th
dimension has index $p_{i,2n}=(4n+1)p_{i,0}-2n$, and for an odd number of steps
it has index $p_{i,2n+1}=-(4n+3)p_{i,0}+2n+2$,
$n=0,1,2,\ldots$. The amplitudes satisfy (\ref{eq13}), but with
\begin{equation}\label{eq16}
\sqrt{\lambda_d^{(i)}} =
\frac{|\mu|A_{\boldsymbol{q}_0}^2}{2^{4-d}\pi
|q_{i,0}-N/2|}=\frac{|\mu| b_{\boldsymbol{k}_0}}{2^{4-d}
|k_{i,0}-\pi/2+\pi/(2N+2)|}.
\end{equation}
For large enough $N$ equation (\ref{eq16}) approaches
the expression (\ref{eq14}), therefore we will use
(\ref{eq14}) for large $N$ only.

\section{$q$-breathers for $d=1$: results}
\label{results1d}

\subsection{Close to the band edge}

The key property of $q$-breathers is that they are localized in
the space of linear normal modes. Note that some $q$-breather
solutions may be compact in $q$-space and contain only one seed mode
$q_0$ \cite{SPO}, or a few modes additionally, due to symmetries of the
interaction network spanned by the nonlinear terms \cite{bushes}.
\begin{figure}[h]
{\begin{center}
\resizebox*{0.70\columnwidth}{!}{\includegraphics{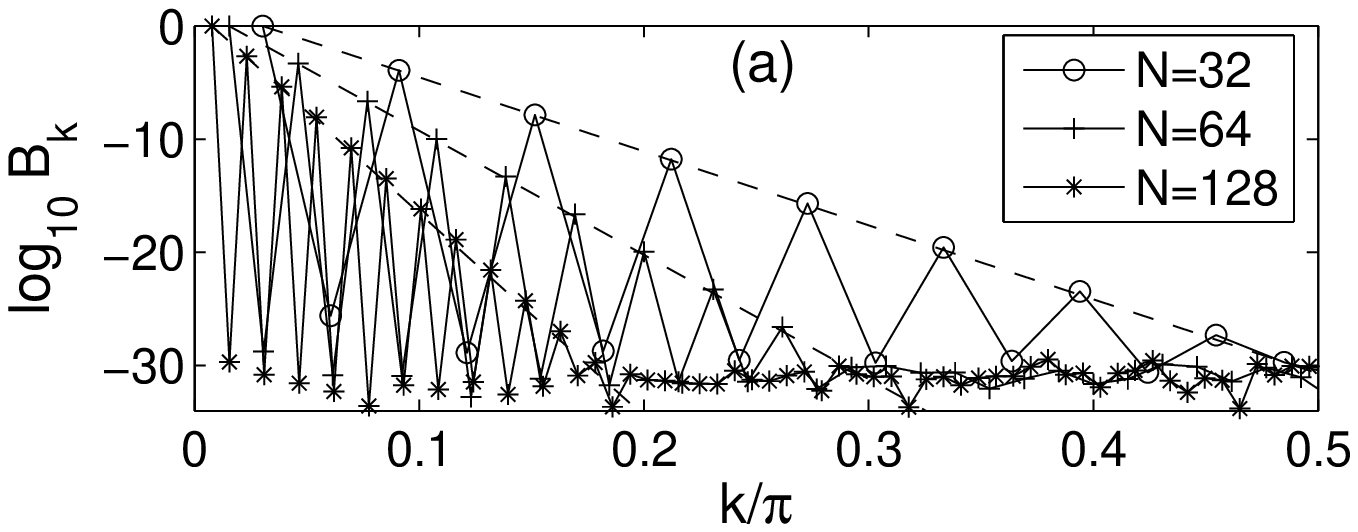}}
\resizebox*{0.70\columnwidth}{!}{\includegraphics{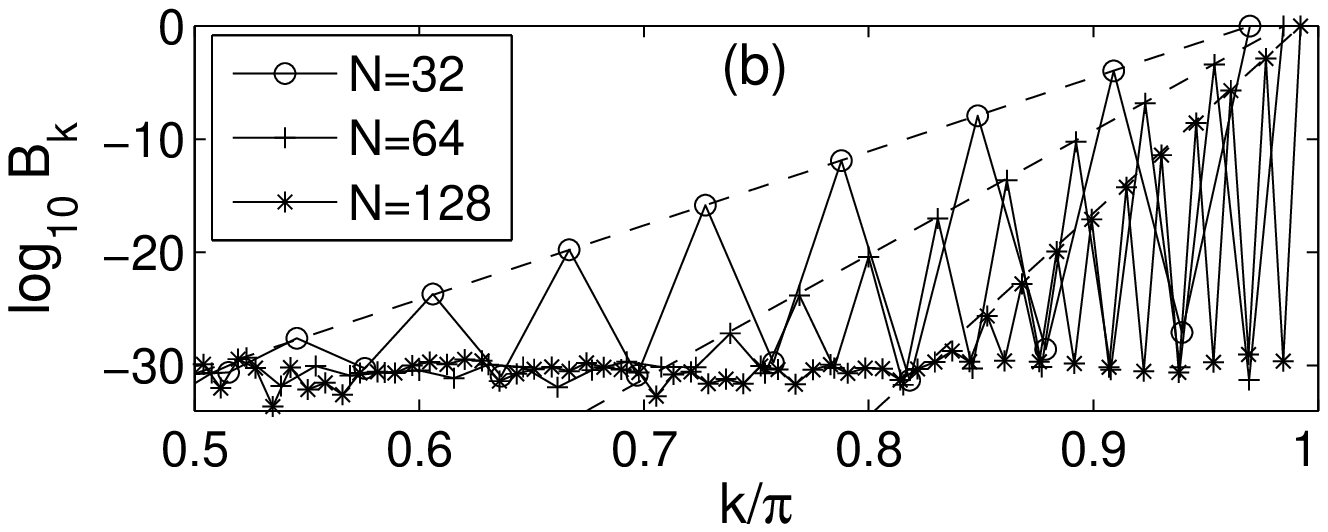}}
\resizebox*{0.70\columnwidth}{!}{\includegraphics{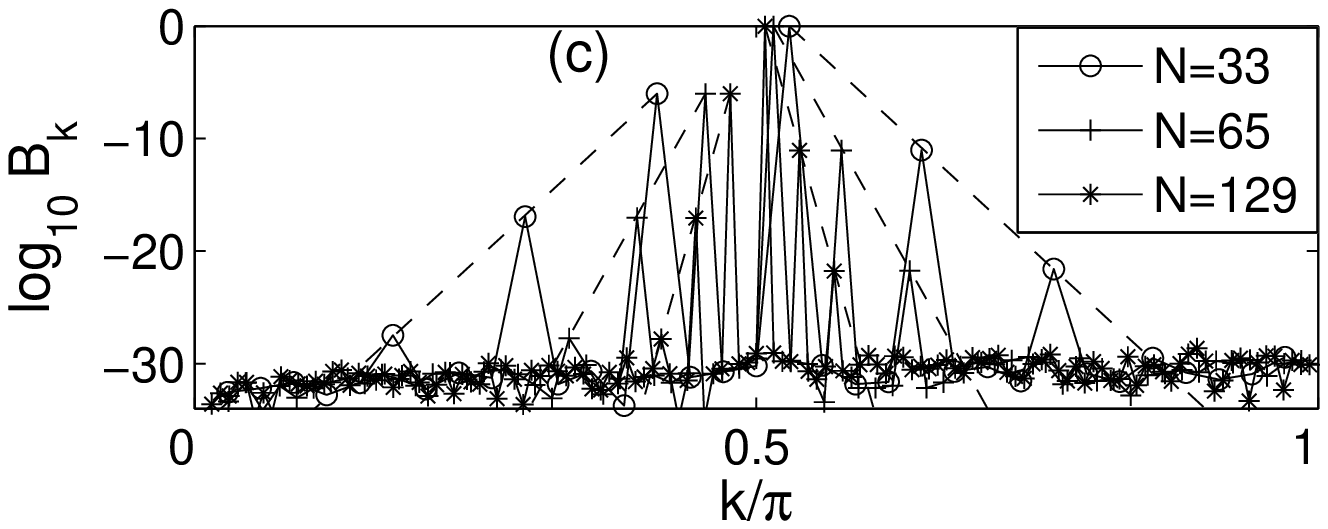}}
\resizebox*{0.70\columnwidth}{!}{\includegraphics{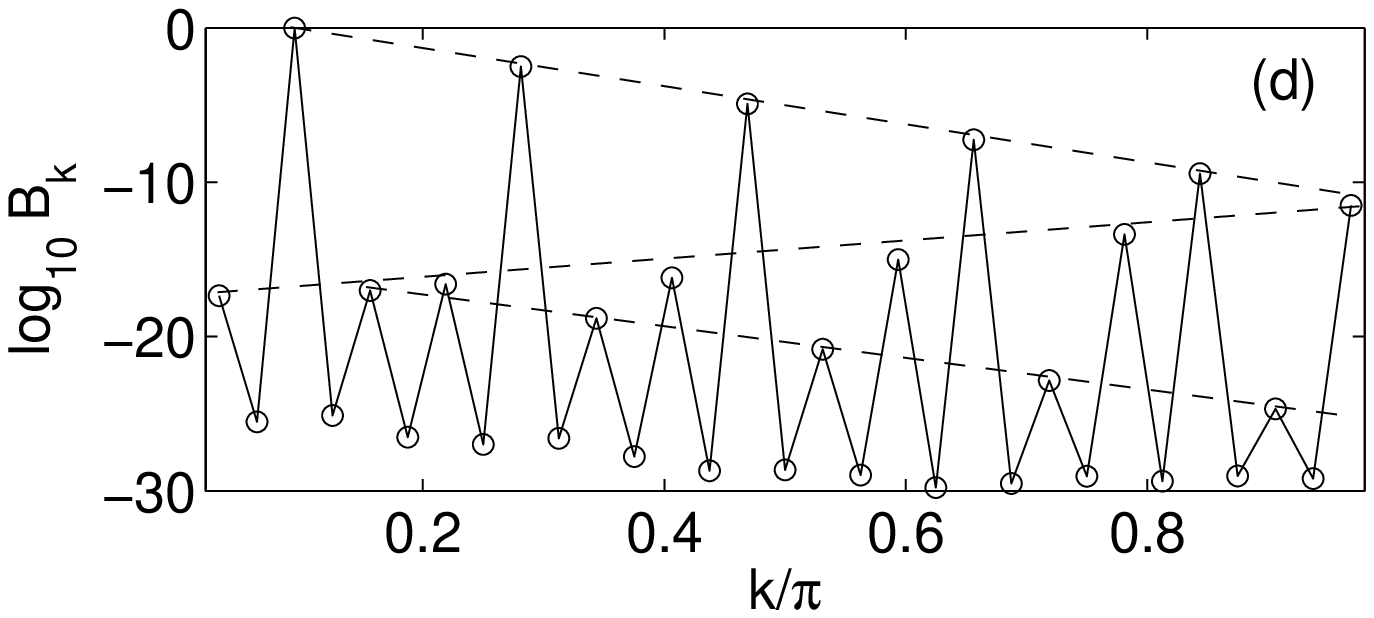}}
\end{center}}
{\caption{Distributions of mode norms for QBs in the
one-dimensional DNLS model with parameters: $B=1$, $\mu=0.1$ for
(a), (b), (c); (a)  QBs with low-frequency seed mode $q_0=1$, (b)
QBs with high-frequency seed mode $q_0=N$, (c) QBs with seed mode
near the middle of the spectrum $q_0=1+(N+1)/2$ (c). Dashed lines
in (a), (b), (c) correspond to analytical estimations. (d)
Observation of multiple reflections at the boundaries of the
$q$-space, $N=31$, $B=1$, $\mu=2$, $q_0=3$. Dashed lines in (d)
are guidelines for the eye.}\label{fig1}}
\end{figure}
\clearpage

Let us consider the one-dimensional case. We compute
$q$-breathers as the stationary solutions of the nonlinear
equations (\ref{eq6}) using the single-mode solution for $\mu=0$ as an
initial approximation. Fig. \ref{fig1}a,b shows the distribution of
mode norms for $q$-breathers in the space of wave numbers $k=\pi
q/(N+1)$ for different chain sizes and
different seed wave numbers, located near the lower ($k_0<<\pi$, Fig.
\ref{fig1}a) and upper ($\pi-k_0<<\pi$, Fig. \ref{fig1}b) edges of
the linear mode spectrum.

We find exponential localization of $q$-breathers, with
a localization length which depends strongly on the chosen parameters.
The obtained analytical estimations for $q$-breathers localized
at the lower band edge (\ref{solutionloweredge}),(\ref{eq8}) respectively upper
band edge (\ref{solutionupperedge}),(\ref{eq8_2}),
are in quantitative agreement with
numerical results (see dashed lines in Fig. \ref{fig1}a,b).

\begin{figure}[t]
{\begin{center}
  \resizebox*{0.90\columnwidth}{!}{\includegraphics{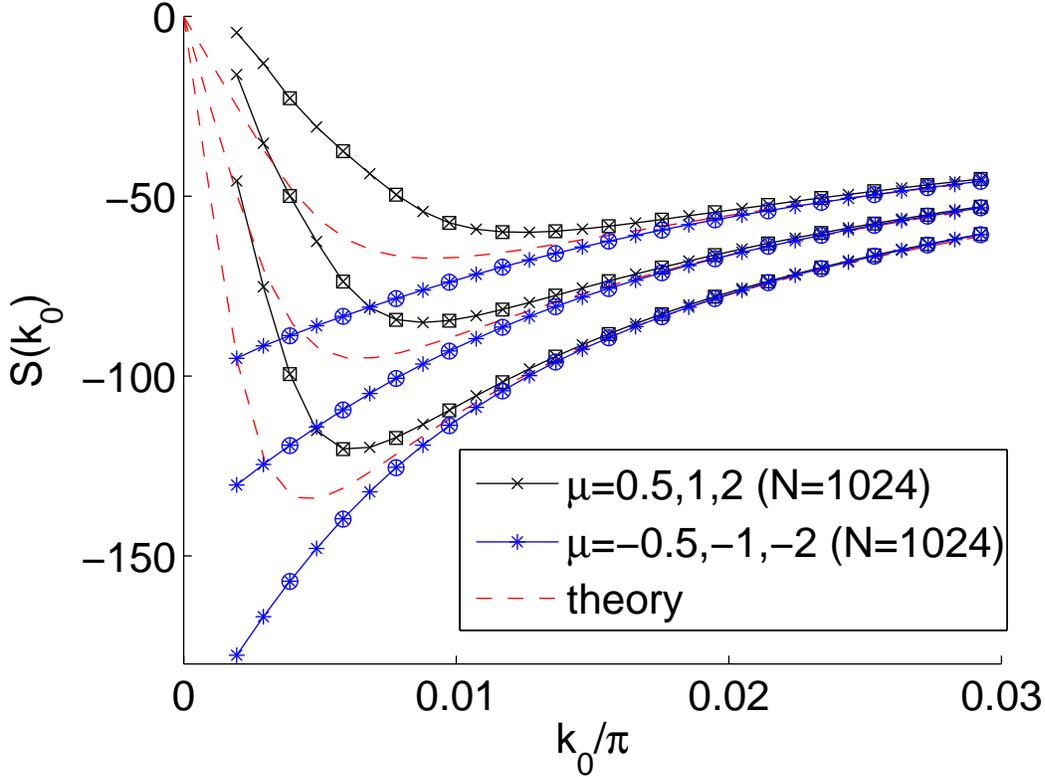}}
 \end{center}}
 {\caption{The slope $S$ as a function of the seed wave
 number $k_0$ for $d=1$ with fixed norm density
 $b=1/1025$, $\mu=\pm0.5,\pm1,\pm2$ (from bottom to top). Dashed lines
 correspond to the analytical  estimates. Different symbols:
 slope from the numerical calculation of QB (squares and circles
 represent the results for $N=512$).
 Solid lines guide the eye.}\label{fig2}}
\end{figure}
The exponential decay in (\ref{eq8}) is depending on the seed mode
norm density. Many applications (e.g. cold atoms in a condensate)
rather fix the total norm, or total norm density. For the obtained
$q$-breather solutions with $q_0<<N$, the relation between these
quantities can be estimated by the sum of infinite geometric
series:
\begin{equation}\label{eq9}
 B\approx\sum\limits_{n=0}^{\infty}
 \lambda^n B_{q_0} = \frac{B_{q_0}}{1-\lambda},
\end{equation}

where $\lambda\equiv\lambda_1^{(1)}$. From
(\ref{eq9}) it follows that $b_{k_0}=(1-\lambda)b$, where $b=B/(N+1)$
is the total norm density. Substituting the expression for
$b_{k_0}$ into (\ref{eq8}) and solving the equation for
$\sqrt{\lambda}$ we obtain:
\begin{equation}\label{eq10}
 \begin{array}{cc}
  \sqrt{\lambda}=\frac{\sqrt{1+4\nu^4/k_0^4}-1}{2\nu^2/k_0^2}, &
  \nu^2=\frac{|\mu| b}{16}.
 \end{array}
\end{equation}
The same dependence of $\sqrt{\lambda}$ on $k_0$
was obtained for low frequency $q$-breathers in the $\beta$-FPU model
\cite{Scaling}.
The exponential
decay of mode norms in the space of wave numbers, can be now
written for $q$-breathers localized in low-frequency
modes:
\begin{equation}\label{eq11}
 \ln b_k=\left(\frac{k}{k_0}-1\right)\ln\sqrt{\lambda}+\ln b_{k_0}.
\end{equation}
To characterize the degree of localization in $k$-space,
we use the
slope of the profile of mode norms in log-normal plots (\ref{eq11}) -- $S$
\cite{Scaling}, where the absolute value of $S$ is equivalent to the inverse
localization length $\xi$:
\begin{equation}\label{eq12}
 S=\frac{1}{k_0}\ln\sqrt{\lambda}\;,\; |S|\equiv \xi^{-1}\;.
\end{equation}
Substituting the expression for $\sqrt{\lambda}$ (\ref{eq10}) into
(\ref{eq12}) we
obtain:
\begin{equation}\label{eq12-final}
 S=\frac{1}{\nu z}\ln\left( \sqrt{1+z^4/4}-z^2/2 \right)\;,\;z=k_0/\nu\;.
\end{equation}
Therefore the slope (inverse localization length) is a function of
the rescaled wavenumber $z$. It therefore parametrically depends on just one
effective nonlinearity parameter $\nu$, which is given by the product of the
total
norm density and the absolute value of nonlinearity strength.
$S$ vanishes for $z\rightarrow
0$, and it has an extremum $\min(S)\approx -0.7432/\nu$ at $z_{min}
\approx 2.577$. The wave number $k_0=k_{min}\equiv z_{min}\nu$
corresponds to the strongest localization of a $q$-breather with
fixed effective nonlinear parameter $\nu$. With increasing $\nu$
the localization length increases.
Most importantly, the localization length diverges for small $k_0 \ll k_{min}$
since $|S| \approx z/(2\nu)$ in that case. This delocalization is due to
resonances with nearby normal modes close to the band edge.
Note, that our analytical
estimations do not depend on the sign of the nonlinearity parameter
$\mu$.

In Fig. \ref{fig1}a,b, due to the small sizes of the chain, all
values of $k_0$ are greater than $k_{min}$, so a monotonous
dependence of the slope on $k_0$ is observed. In Fig. \ref{fig2}
we plot theoretical and numerically obtained dependencies $S(k_0)$
for different values of the nonlinear
parameter $\mu$ and different system sizes $N$ (we use large
enough $N$ to resolve the theoretically predicted extremum of $S$).
In all cases the numerical results show, that the slopes indeed are characterized
by intensive quantities only, and the above derived scaling laws hold.

For positive values of $\mu$ and seed wave numbers close to the lower band edge,
the extremum in $S$ is reproduced in the numerical data,
though the numerical curves
deviate from theoretical curves for small $k_0$. We also find that
$k_{min}$ increases and $|S(k_{min})|$ decreases with increasing
$\mu$ as it is predicted by our analytical results (increase of
norm density $b$ gives the same effect).

\begin{figure}[h]
{\begin{center}
 \resizebox*{0.70\columnwidth}{!}{\includegraphics{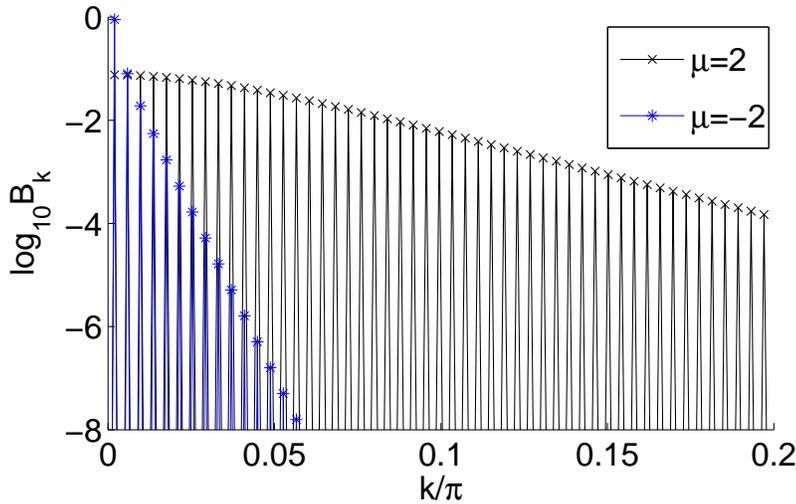}}
 \end{center}}
 {\caption{Distributions of mode norms for QBs in the one-dimensional DNLS
 model for positive and negative values of $\mu$;
 $q_0=2$, $N=1024$.}\label{fig3}}
\end{figure}

For negative values
of $\mu$ and seed wave numbers close to the lower band edge,
we do not observe an extremum for $S$. In the region of
small $k_0$ the numerically obtained curves for positive and negative
values of $\mu$ differ, while the theoretically predicted slopes
do not depend on the sign of $\mu$.
Fig. \ref{fig3} shows the mode norm profiles
of $q$-breather solutions with small $k_0$ obtained for positive
and negative values of $\mu$.
The reason for the discrepancy between theoretical prediction
and numerical results for negative values of $\mu$
must be strong contributions from higher order terms in the
perturbation expansion.
The standard argument is, that the perturbation theory is valid if the localization
length is small, i.e. $|\lambda| \ll 1$ as well.
When $|\lambda|$ becomes of the order of one, higher order terms in perturbation
theory have to be taken into account, and it would be tempting to conclude
that delocalization will take place. That should be true especially
when all higher order terms in the series carry the same sign.
That is the case for positive nonlinearity here, but for negative $\mu$ we
obtain alternating signs of higher order terms. These alternating signs
therefore effectively cancel most of the terms in the series,
and are responsible for an increasing localization of $q$-breathers with negative $\mu$
in the limit of small wavenumbers.

\begin{figure}[h]
{\begin{center}
 \resizebox*{0.90\columnwidth}{!}{\includegraphics{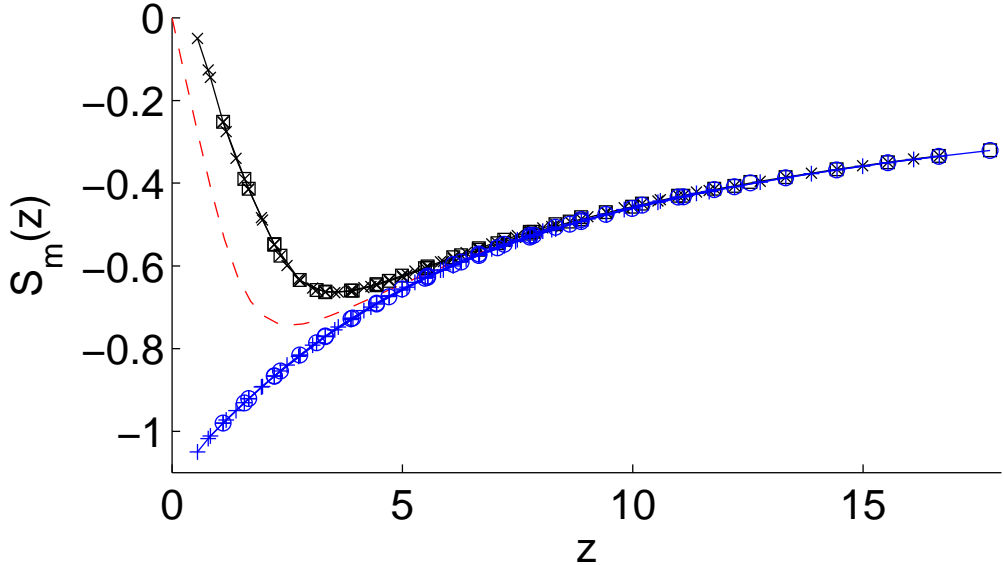}}
 \end{center}}
 {\caption{The master slope function $S_m(z)$ (dashed line). Different
 symbols and eye-guiding solid lines correspond to the scaled numerical
 estimates of the slope presented in Fig. \ref{fig2} (bottom line
 ---  $\mu<0$, top line --- $\mu>0$).}\label{fig4}}
\end{figure}
The obtained results for $q$-breathers with seed wave numbers close to the
lower band edge ($k_0<<\pi$) are valid for the case of $k_0$ close
to the upper band edge ($\pi-k_0<<\pi$) if we change $\mu
\rightarrow -\mu$. Thus, there is a strong asymmetry in the localization
properties of $q$-breather solutions with $k_0<\pi/2$ and
$k_0>\pi/2$ for a fixed sign of nonlinearity.

The numerical results in Fig. \ref{fig2} show that slope values
calculated for different system sizes lie on the same curves with
corresponding $\mu$ even in the region of small $k_0$, where
higher order corrections to our analytical estimates have to be
taken into account. This result is in agreement with the exact
scaling of $q$-breather solutions described in \cite{Scaling}. We
plot in Fig. \ref{fig4} the master slope function $S_m(z)=\nu S$,
which depends on a single variable $z$ \cite{Scaling}. It implies
that knowing this single master slope function is sufficient to
predict the localization property of a $q$-breather at any seed
wave number $k_0\ll\pi$, at any energy etc. Numerically obtained
slopes presented in Fig. \ref{fig2} are rescaled and plotted in
Fig. \ref{fig4}. We see, that all results corresponding to the
same sign of $\mu$ condense on a single curve even (and
especially) for small $k_0$, though these numerical results differ
from the analytical estimation.

\subsection{Close to the center of the band}

Fig. \ref{fig1}c shows the distribution of
mode norms for $q$-breathers in the space of wave numbers $k=\pi
q/(N+1)$ for different chain sizes and
different seed wave numbers, located close to the center of the spectrum
($|k_0-\pi/2|<<\pi$).
The analytical estimation for the amplitudes of these
$q$-breathers  are in good quantitative
agreement with the numerical results, for small enough parameters $\mu$
and $b$, cf. Fig. \ref{fig1}c.

We express again the norm density of the seed mode
via the total norm density: $b_{k_0}=(1-\lambda)b$,
($\lambda\equiv\lambda_1^{(1)}$). Substituting this expression
into (\ref{eq14}) we find:
\begin{equation}\label{eq17}
 \begin{array}{cc}
  \sqrt{\lambda}=\frac{\sqrt{1+4\nu^2/(k_0-\pi/2)^2}-1}{2\nu/|k_0-\pi/2|}, &
  \nu=\frac{|\mu| b}{8}.
 \end{array}
\end{equation}
The slope of the mode norm profile
in $k$-space is given by
\begin{equation}\label{eq18}
 S=\frac{1}{|k_0-\pi/2|}\ln\sqrt{\lambda}=\frac{1}{2\nu
 z}\ln\left( \sqrt{1+z^2}-z \right),
\end{equation}
where $z=|k_0-\pi/2|/(2\nu)$. For $z<<1$:
$S=(-1+z^2/6+O(z^4))/(2\nu)$, for $z>>1$:
$S=\ln(1/(2z)+O(1/z^3))/(2\nu z)$. In the limit $z\rightarrow0$,
the slope $S\rightarrow -1/(2\nu)$.
Thus, the strongest localization of a $q$-breather with fixed
effective nonlinear parameter $\nu$ should be obtained for wave numbers
$k_0\approx \pi/2$.
The increase of the effective
nonlinearity parameter $\nu\sim|\mu|b$ leads to a weaker localization
of $q$-breathers in $k$-space, since the absolute value of the slope $S$
decreases.
Note, that there is a special point for odd
$N$: $k_0=\pi/2$, $q_0=(N+1)/2$. For this seed mode, the $q$-breather
is compact in $k$-space \cite{SPO}.

\begin{figure}[h]
{\begin{center}
  \resizebox*{0.90\columnwidth}{!}{\includegraphics{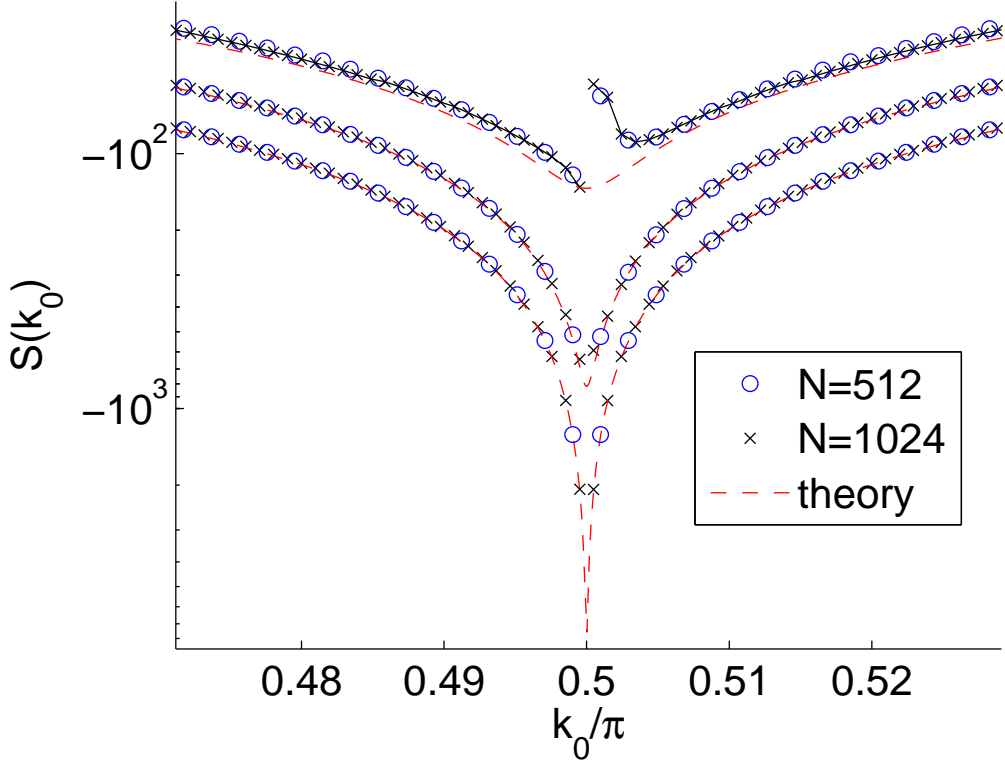}}
 \end{center}}
 {\caption{The slope $S$ as a function of the seed wave number $k_0$
 for $d=1$ with fixed norm density $b=1/1025$, $\mu=0.5,5,30$
 (from bottom to top). Dashed lines
 correspond to the analytical estimates. Different symbols correspond to the
 estimates of the slopes from numerical calculations of QBs.}\label{fig5}}
\end{figure}

In Fig. \ref{fig5} the theoretical and numerically obtained
dependencies $S(k_0)$ for
different values of $\mu$ and different system
sizes $N$ are plotted. For small values of $\mu$ we
observe good agreement between analytical and numerical results. But
the increase of the nonlinearity leads to a deviation between the theoretical
and numerical curves in the region of $k_0$ close to $\pi/2$.
These corrections, as it was for the case of
$q$-breathers localized near the band edges, depend on the location of
$k_0$ ($k_0>\pi/2$ or $k_0<\pi/2$) and on the sign of nonlinearity.
Therefore  the curves of $S(k_0)$ for strong
nonlinearity ($\mu=30$) in Fig. \ref{fig5} are non-symmetric around the
point $k_0=\pi/2$: in contrast to $k_0<\pi/2$, for $k_0>\pi/2$ a
local minimum of $S$ is observed. Still, the predicted scaling properties
of $q$-breathers remain correct even for strong nonlinearity: the values of
$S$, computed for different system sizes $N$, lie on the same
curves for fixed $\mu$.

\subsection{Stability of $q$-breathers}

We analyze the linear stability of $q$-breathers as stationary
solutions of DNLS model considering the evolution of small
perturbations $\varepsilon_n$ in the rotating frame of the
periodic solution \cite{Gorbach}:
$\psi_{\boldsymbol{n}}(t)=(\phi_{\boldsymbol{n}}^0+\varepsilon_{\boldsymbol{n}}(t))
\exp(i\Omega t)$, where $\phi_{\boldsymbol{n}}^0$ are the
non-perturbed time-independent amplitudes,
$\varepsilon_{\boldsymbol{n}}=\alpha_{\boldsymbol{n}}+i\beta_{\boldsymbol{n}}$.
A periodic orbit is stable when all perturbations do not grow in
time. Solving the linearized equations for the perturbation,
that condition translates into the request,
that all eigenvalues $s_m$ ($m=\overline{1,2N^d}$) of
the linearized equations must be purely imaginary.
Otherwise the orbit is unstable.
\begin{figure}[h]
{\begin{center}
\resizebox*{0.8\columnwidth}{!}{\includegraphics{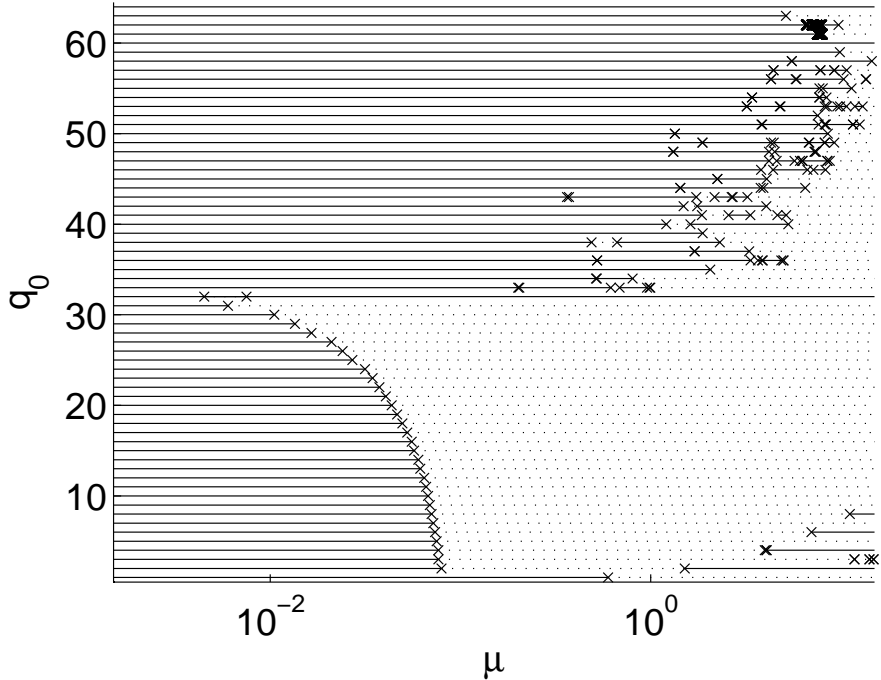}}
\end{center}}
{\caption{Domain of stability of $q$-breathers
for $d=1$. Solid lines - stable, dashed - unstable, crosses - switch from
stable to unstable or vice versa.
$N=64$,
$B=2$.}\label{fig8}}
\end{figure}
In Fig. \ref{fig8} we plot the numerical outcome of the stability analysis. We smoothly
continue
$q$-breather solutions for each seed mode number $q_0$ by
increasing the nonlinearity parameter $\mu$ and check the stability.
If the maximum absolute value of the real parts of all eigenvalues
is smaller than $10^{-6}$, the $q$-breather is considered as stable
(solid line), otherwise it is unstable (dashed line), crosses
mark the change of stability. For small values
of nonlinearity all $q$-breathers are stable. Qualitatively
different threshold values and dependencies on $q_0$ for
$q$-breathers with seed modes from different parts of the spectrum
$q_0<N/2$ and $q_0>N/2$ are observed. This is in agreement
with the results presented in \cite{SW}, where stability
properties of nonlinear standing waves are studied.
Note, that due to the symmetry of the equations,
the stability properties of a $q$-breather with seed mode $q_0$ for
some negative value of the nonlinearity parameter $\mu$ is the same as the
stability property of the $q$-breather with seed mode
$\widetilde{q}_0=N+1-q$ for the nonlinearity parameter $\widetilde{\mu}=-\mu$.

Let us discuss the possible link between linear stability and localization.
If a $q$-breather becomes delocalized, that happens because of resonances
between different mode frequencies. Therefore we can expect, that the same
resonances
will also drive the state unstable. Indeed, these correlations can be clearly
observed from the numerical data. However, if a $q$-breather is well localized,
it does not follow that it will be stable as well, since instability can arise
due to resonant interaction of modes in the breather core alone.

\section{Periodic boundary conditions}
\label{pbc}

In the case of periodic boundary conditions, we have used the following
transformations between real space and the reciprocal space of
normal modes (for even $N$):
\begin{equation}\label{eq20}
 \begin{array}{l}

  \psi_{\boldsymbol{n}}(t)=\frac{1}{N^{d/2}}\sum\limits_{q_{1},\ldots,
  q_{d}=-N/2+1}^{N/2} Q_{\boldsymbol{q}}(t)\prod\limits_{i=1}^d
  \exp{\left(\frac{2\pi q_i (n_i - 1)}{N}\right)}\;.
 \end{array}
\end{equation}
The DNLS model (\ref{eq1}) with periodic boundary conditions has
exact solutions for nonlinear traveling waves, which can be
written, for instance in case $d=1$, as: $\psi_n(t)=\phi_0
\exp(i\Omega t - i k_0 n)$, where $\Omega=-2\cos k_0
-\mu\phi_0^2$, $k_0=2\pi q_0/N$, $q_0\in [-N/2,N/2]$. These types
of solutions can be considered as compact $q$-breathers which
contain only one mode $q_0$. Traveling modes are also not
invariant under time reversal.
\begin{figure}[t]
{\begin{center}
  \resizebox*{0.90\columnwidth}{!}{\includegraphics{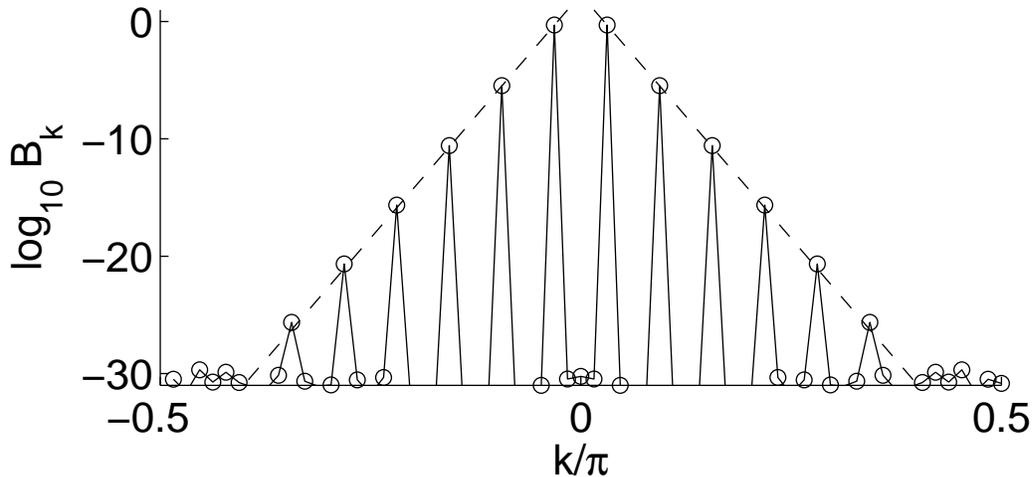}}
 \end{center}}
{\caption{Time-reversible $q$-breather in the one-dimensional DNLS
model with periodic boundary conditions, continued from the linear
standing wave consisting of two traveling waves with $q_0=2$ and
$q_0=-2$, $N=64$, $B=1$, $\mu=0.1$. Dashed lines represent
analytical estimations.}\label{fig6}}
\end{figure}
The
continuation of a linear standing
wave, consisting of two traveling waves with the same norms and
wave numbers: $k_0$ and $-k_0$, into the nonlinear regime leads to
a time-reversible $q$-breather solution (see Fig. \ref{fig6}), which
is not compact, and its localization properties are similar to the
properties of $q$-breathers in the case of fixed boundary
conditions. Here we present the result for decay of mode
norms $\lambda_d^{(i)}$ for the case $|k_{i,0}|<<\pi$, which
differs from (\ref{eq8}) by a prefactor:
\begin{equation}\label{eq19}
 \begin{array}{cc}
  \sqrt{(\lambda_d^{(i)})}=\frac{|\mu| A_{\boldsymbol{q}_0}^2 N^{2-d}}
  {32 \pi^2{(q_{i,0})}^2} = \frac{|\mu| b_{\boldsymbol{k}_0}}
  {8 {(k_{i,0})}^2}\;, &
  i=\overline{1,d}\;,
 \end{array}
\end{equation}
where $b_{\boldsymbol{k}_0}=B_{\boldsymbol{k}_0}/N^d$,
$k_{i,0}=2\pi q_{i,0}/N$. Fig. \ref{fig6} illustrates good quantitative agreement between  
analytical and numerical results obtained for small enough values of norm and nonlinearity.

\section{$q$-breathers in two- and three-dimensional lattices}
\label{results2d3d}

\begin{figure}[h]
{\begin{center}
\resizebox*{0.49\columnwidth}{!}{\includegraphics{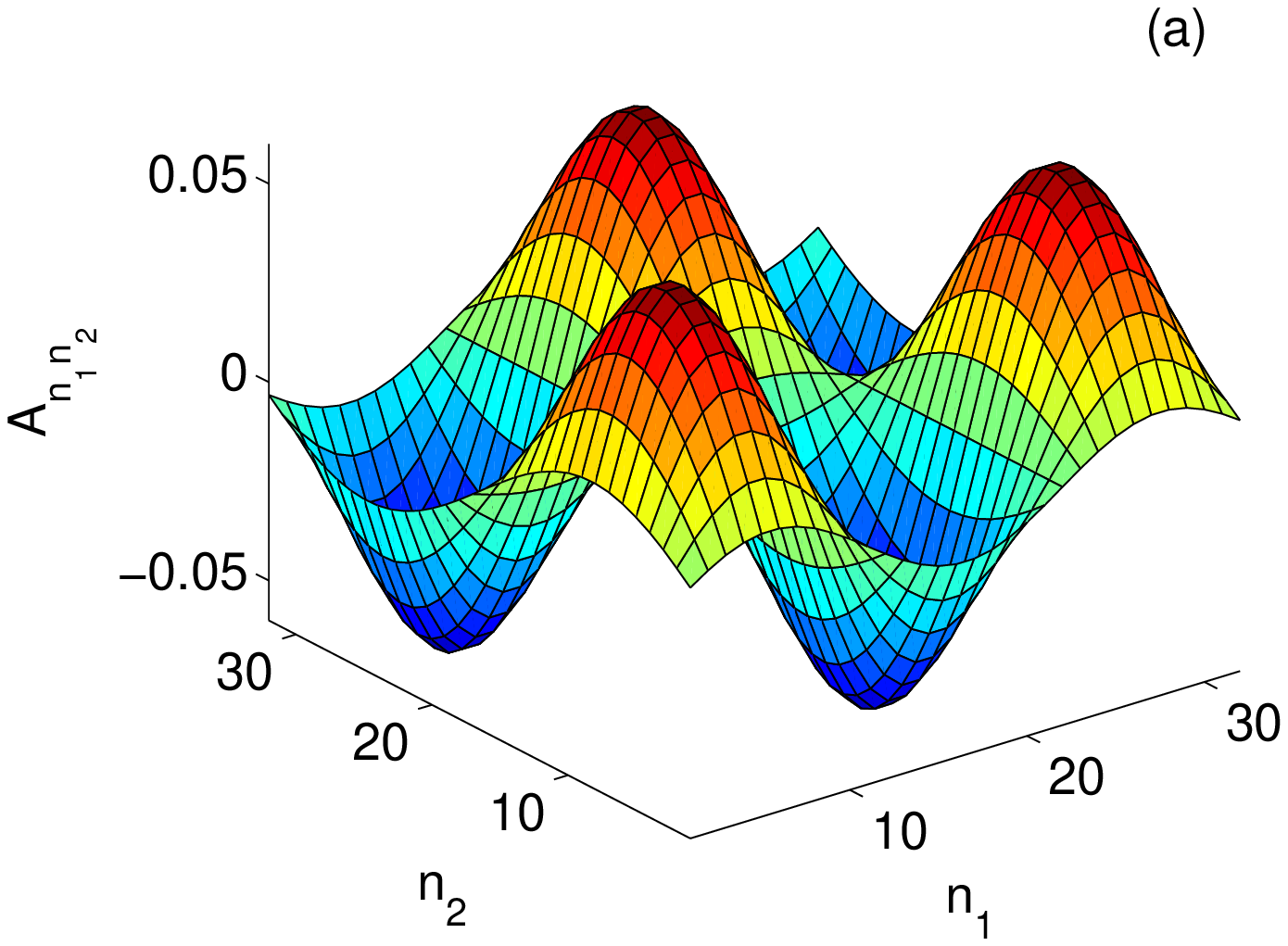}}
\resizebox*{0.49\columnwidth}{!}{\includegraphics{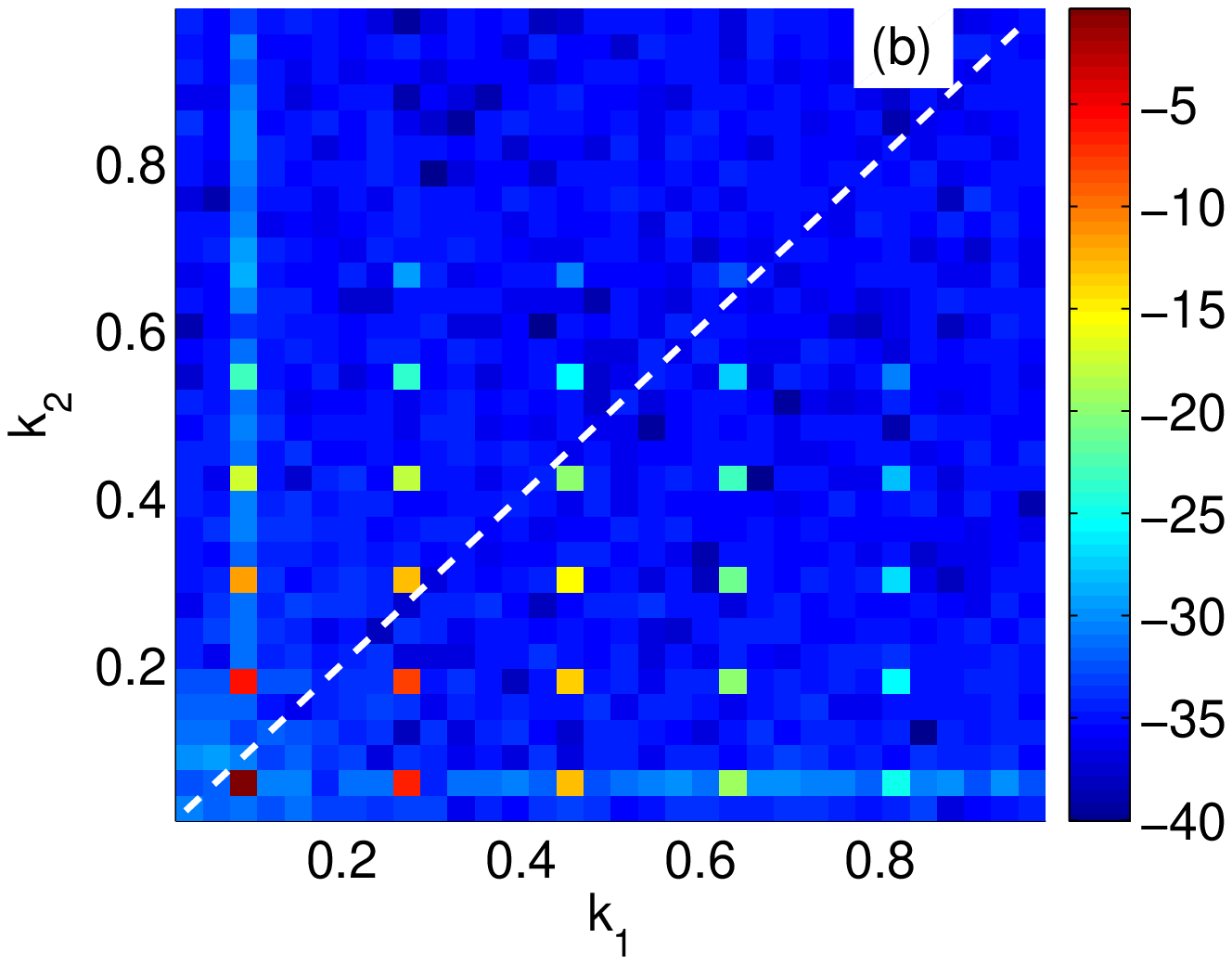}}
\resizebox*{0.49\columnwidth}{!}{\includegraphics{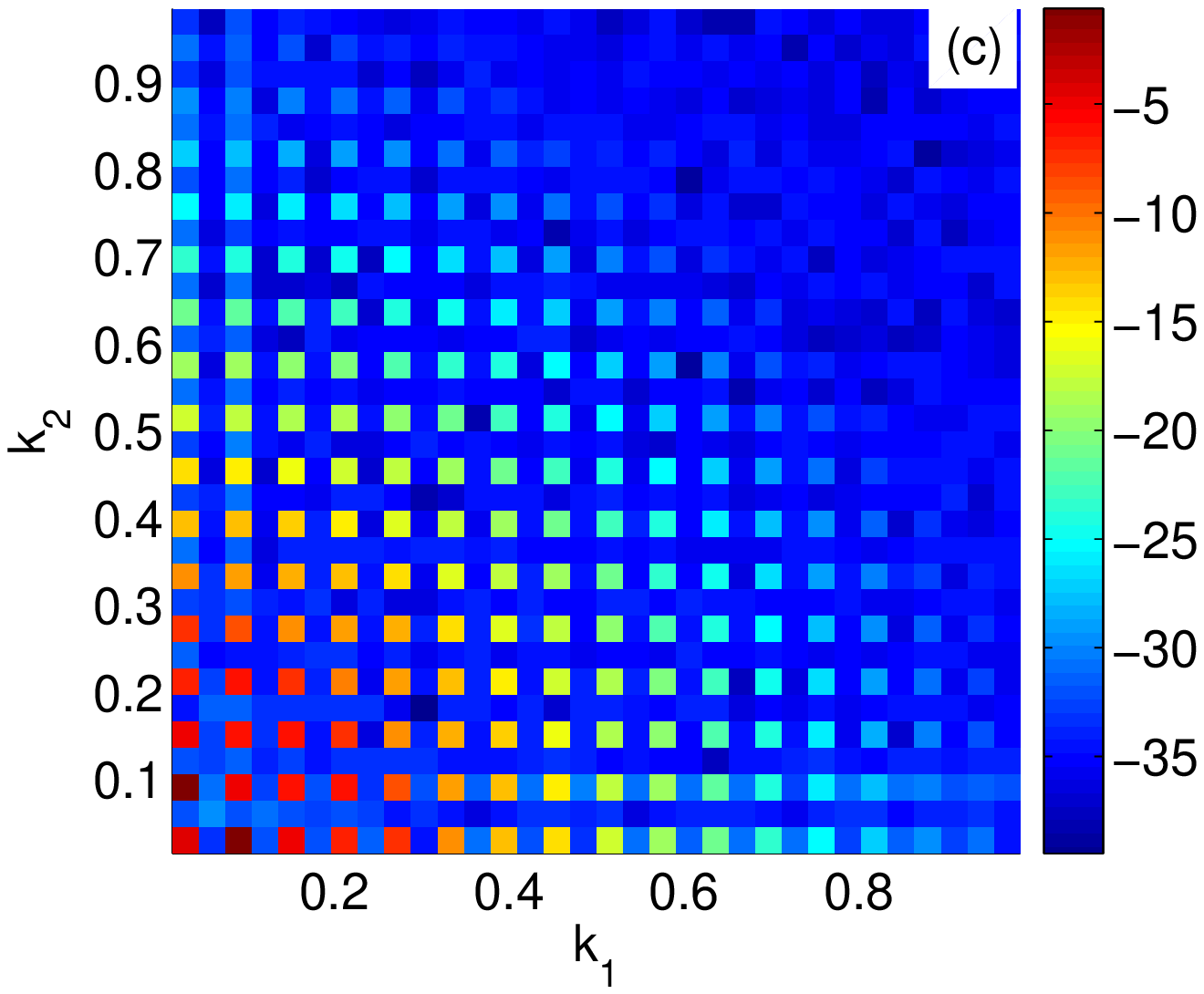}}
\resizebox*{0.49\columnwidth}{!}{\includegraphics{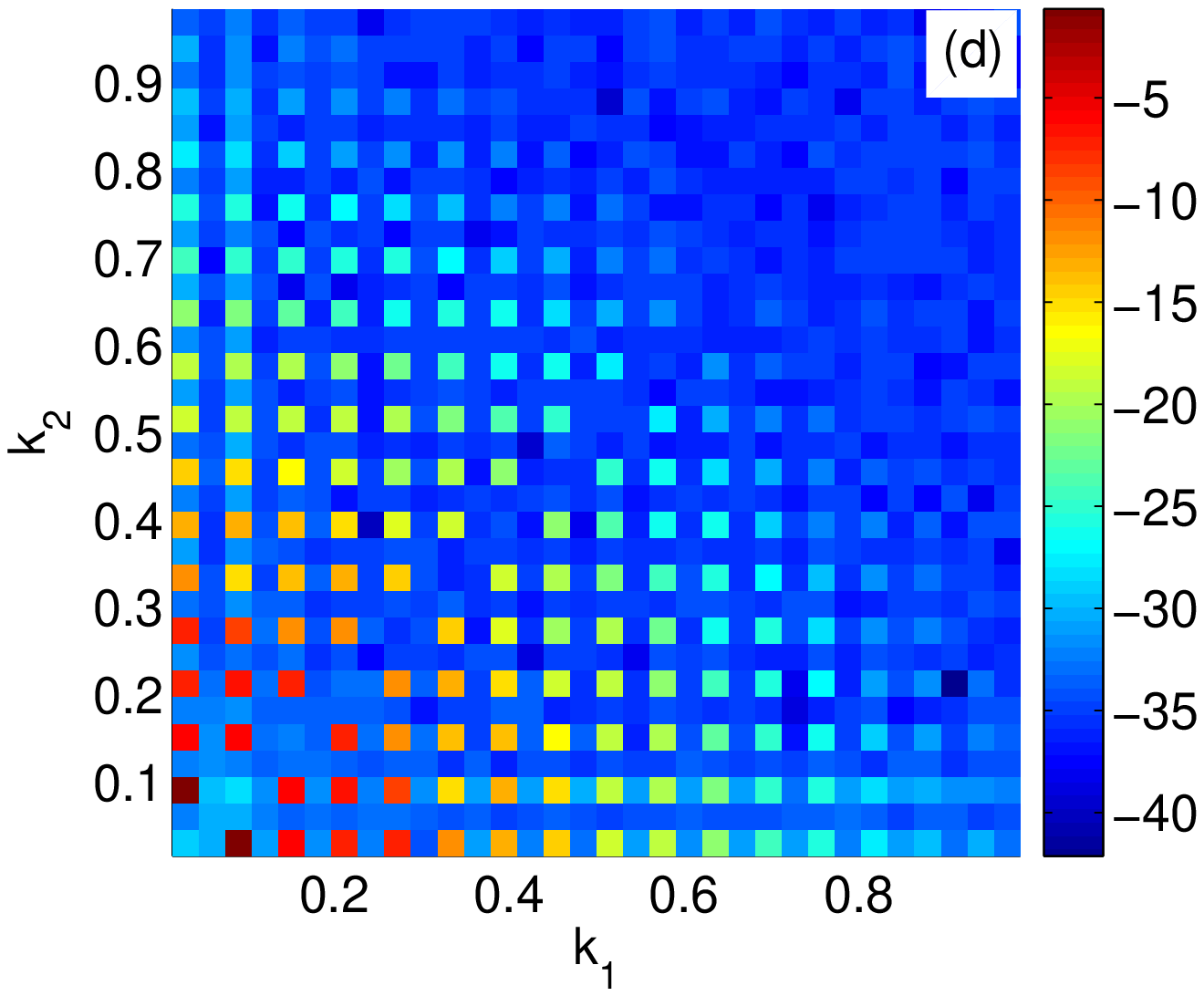}}
\end{center}}
{\caption{Different $q$-breather modes for $d=2$: $N=32$, $B=1$,
$\mu=0.5$. Mode norm magnitude is plotted in color code on
logarithmic scale (except for (a), where a linear scale is used).
(a), (b) --- asymmetric single-mode $q$-breather with seed mode
($\boldsymbol{q}_0=(3,2)$) in real space (a) and in $q$-space (b);
(c), (d) --- symmetric multi-mode $q$-breather with in-phase (c)
and antiphase (d) seed mode pair $\boldsymbol{q}=(3,1),(1,3)$.
Dashed line in (b) guides the eye along the main
diagonal.}\label{fig7}}
\end{figure}

For two- and three-dimensional symmetric DNLS lattices (\ref{eq1})
with fixed boundary conditions, only linear modes with mode
numbers $\boldsymbol{q}$ on the main diagonal have non-degenerate
frequencies. Using the Implicit Function Theorem \cite{ImpFunc},
it follows that these modes are continued into the nonlinear
regime. However, we also successfully performed numerical continuations
of $q$-breathers with seed mode numbers off the main diagonal as
it was done in \cite{FPU2d3d} for the FPU model. For the
two-dimensional DNLS model the $q$-breather, continued from the
single linear mode $\boldsymbol{q}_0=(2,3)$, is presented in Fig.
\ref{fig7} in real space (a) and in $q$-space (b). The
$q$-breather with the seed mode $\boldsymbol{q}_0=(3,2)$ exists as
well and has the same frequency. The slowest decay of the mode
norms happens to be along the direction of the main axes. The
decay is exponential and in good agreement with the analytical
estimation (\ref{eq8}) for $d=2$.
In addition to such asymmetric single mode $q$-breathers it is
possible to construct multi-mode $q$-breathers continued from a
pair of degenerate linear normal modes $(q_{0},p_{0})$ and
$(p_{0},q_{0})$ ($q_{0}\neq p_{0}$) with the same norms in both
modes and in-phase (Fig. \ref{fig7}c) or antiphase (Fig.
\ref{fig7}d) oscillations. It is important to note that the
problem of degenerate frequencies is avoided for these solutions.
Indeed, system (\ref{eq4}) has two invariant manifolds
$Q_{q_1,q_2}=\pm Q_{q_2,q_1}$. Looking for a solution on a
manifold, the number of independent variables of state is reduced
from $N^2$ to the dimensionality of the manifold, which equals
$(N^2+N)/2$ for the symmetric manifold and $(N^2-N)/2$ for the
antisymmetric one. The reduced system of equations contains only modes
with non-degenerate frequencies.

For $d=3$ we have also verified, that the analytical estimations of mode
norm decay (\ref{eq8}), (\ref{eq14}) agree well with
the results of
numerical calculations for single-mode $q$-breathers.

\begin{figure}[h]
{\begin{center}
\resizebox*{0.9\columnwidth}{!}{\includegraphics{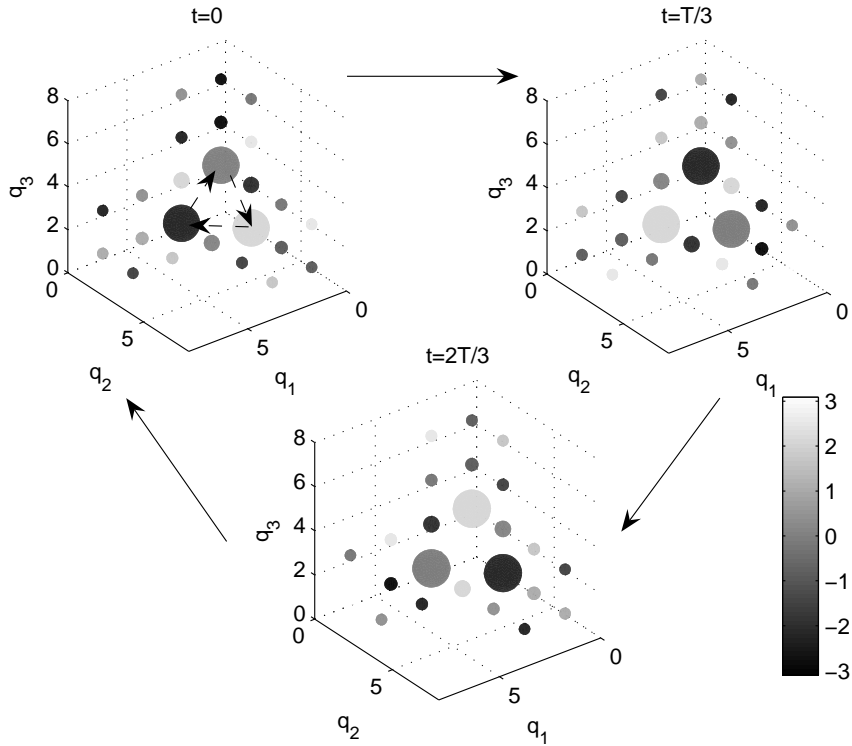}}
\end{center}}
{\caption{A vortex $q$-breather state for $d=3$: $N=8$, $B=1$,
$\mu=0.1$, at some fixed time. Mode norm magnitude is encoded in
sphere sizes on logarithmic scale, while the sphere color denotes
the phase of the mode. Seed mode triplet: $\boldsymbol{q}=(3,1,1),
(1,3,1), (1,1,3)$.}\label{fig9}}
\end{figure}

In addition to various time-reversible $q$-breather solutions,
which are constructed in the same way as for $d=2$, the
three-dimensional DNLS model allows also for non-time-reversible
(``vortex'') multi-mode solutions. Let us consider $q$-breather
solutions on an invariant manifold of the system (\ref{eq4}):
$Q_{q_1,q_2,q_3}=\exp(2\pi i/3) Q_{q_3,q_1,q_2} = \exp(4\pi i/3)
Q_{q_2,q_3,q_1}$ (note, that this manifold has a counterpart with
the opposite sign of the phase shifts). On the manifold the number
of variables of state is reduced to $(N^3-N)/3$. We have
constructed numerically a vortex $q$-breather solution continued
from a degenerate triplet of seed modes which have the same norm
and a relative phase shift $2\pi/3$: $Q_{q_0,q_0,p_0}=\exp(2\pi
i/3)Q_{p_0,q_0,q_0}=\exp(4\pi i/3)Q_{q_0,p_0,q_0}$, $q_0\neq p_0$.
The frequency of such a triplet becomes non-degenerate in the
reduced system on the manifold. The energy flows in a vortex-like
manner in $q$-space for such excitations (Fig.\ref{fig9}).

\section{Discussion}
\label{discussion}

Comparing the aims we formulated in the introduction with the above results,
we can conclude, that indeed the $q$-breather concept turns out to be generic,
and time-periodic orbits, which are localized in normal mode space, are probably
as generic as discrete breathers (although perhaps in a different way).
Along with the study of the details of $q$-breathers properties in DNLS models,
we found some particularities, which were not (yet) obtained in acoustic (FPU)
systems. Let us discuss some of these and other findings from above.

\subsection{Scales, delocalization thresholds, and healing length of a BEC}

Let us consider a finite DNLS chain with $N$ sites. Let us fix the
total norm density $b$ and the nonlinearity $\mu$. Therefore we
fix the effective nonlinearity parameter $\nu^2=|\mu|b/16$
(\ref{eq10}). If the system size was large enough, we will resolve
the extremum in $S$, and therefore the QB with seed wave number
$k_{min} \approx 2.577 \nu$ is the strongest localized one. It
evidently sets an inverse length scale $\nu$. This length scale is
known in the GP equation for a BEC in a trap with (similar) fixed
boundary conditions. One has a condensate, whose amplitude
vanishes at some boundary, yet getting back to some mean value
away from the boundary - exactly at the healing length $\xi_h =
(4\pi a n)^{-1/2}$, where $n$ is the condensate density, and $a$
is the scattering length which is proportional to the atom-atom
interaction, and therefore to the nonlinearity strength $\mu$
within the mean field GP equation \cite{healinglength}. Therefore,
the inverse healing length corresponds to the wave number scale,
on which the most strongly localized QB is observed.

We now increase the system size further, and compute the localization length
(or respectively its negative inverse - the slope $S$) of the longest wavelength mode.
With increasing $N$ the grid of allowed $k$-values becomes denser, and at
some critical $N_c(\nu)$ the localization length will reach the finite size of
the normal mode space, and the $q$-breather delocalizes. With a little algebra it follows
\begin{equation}
N_c \approx \frac{\pi^2}{2\nu^2} \;,\;k_0^{(c)} \approx \frac{2\nu^2}{\pi}\;.
\label{discussion1}
\end{equation}
For even larger (and finally infinitely large) lattices the critical value $k_0^{(c)}$
is marking a border in $k$-space: at the given norm density $\nu$, all QBs with seed wave
numbers $k \gg k_0^{(c)}$ are localized, while we obtain delocalization for $ k \le k_0^{(c)}$.
So there is a layer of delocalized QBs at the band edge, whose width grows according to
(\ref{discussion1}) with growing $\nu$. Modes launched inside this layer (with the given
norm density) will quickly spread their energy among many other modes - they will
quickly relax. Modes launched outside this layer will stay localized in normal mode space,
at least for sufficiently long times. $k_0^{(c)}$ is therefore separating a layer of strongly
interacting modes from weakly interacting ones. For large enough effective nonlinearity
parameter $\nu \sim 1$ the whole wave number space is filled with strongly interacting
normal modes, and the normal mode picture breaks down completely.
As long as $\nu$ is smaller, the value of $k_0^{(c)}$ sets a new length scale $2\pi/k_0^{(c)}$,
which is proportional to the {\sl squared} healing length $\xi_h^2$.
On that new length scale, the normal mode picture breaks down.

\subsection{Delocalization thresholds and modulational instability}

It is instructive to remember, that delocalization thresholds of QBs are related to
resonances. Indeed, in the present case, the density of states at the band edge
diverges in the limit of large system size, and many modes have almost the same
frequencies. It is these small differences, which tend to zero as $1/N^2$, and which
are responsible for the resonant mode-mode interaction. Notably the analysis of
stability of band edge modes \cite{bem} (note: for periodic boundary conditions)
shows some interesting correlations.
The analyzed band edge modes are compact in normal mode space, yet they undergo a
tangent bifurcation at amplitudes, which are the smaller, the larger the system size.
These instabilities appear however only for a certain sign of the nonlinearity,
which exactly corresponds to the actual observed delocalization of QBs.
Therefore we expect, that the (so far not studied) case of a FPU chain with negative
quartic nonlinearity, which is known {\sl not} to yield an instability for the band edge mode,
will {\sl not} show delocalization of QBs close to the (upper) band edge.
It is furthermore instructive, that if a tangent bifurcation of a band edge mode takes place,
the simulation of a lattice shows the onset of modulational instability, which
leads to a collection of energy in smaller system volumes, and finally to the formation
of discrete breathers - i.e. to localization in real space. Thus we may expect,
that the resonant layer of strongly interacting modes may lead to the formation
of localized states in real space, while the rest of the normal mode space is evolving
in the regime of localization in normal mode space. Thus, we may expect to observe
in an actual simulation localization {\sl both}  in normal mode space {\sl and} in real space.

\subsection{Comparing analytical and numerical results}

While the delocalization at one band edge, as predicted by
perturbation theory, is reproduced by numerical data, that does
not happen at the second band edge (both band edges change their
places, when the nonlinearity inverts sign). The breakdown of
perturbation theory follows from comparing the leading and
next-to-leading order terms in the expansion. When terms are of
the same order, we conclude, that the series will diverge, and
therefore the QB solutions will delocalize. This is true for the
case when all terms in the series have the same sign. However,
when the terms have alternating sign, the above conclusion must
not be correct. And indeed the numerical data show, that these
sign alternations lead to an effective cancelation, and a final
strong localization of a QB solution. A similar breakdown of
perturbation theory happens for QB solutions localized at the band
center for large nonlinearity (or norm). The perturbation theory
tells, that QB states are well localized from both sides of the
center. Numerical data however show, that this is true only on one
side from the center, while the other side shows a tendency
towards delocalization.

\subsection{Open questions}

Below we list some potentially interesting and important open questions.
First, the influence of different boundary conditions has not been
systematically studied. QBs are extended states in real space, therefore their
spectrum, and the way they interact,
may to some extend be sensitive to the choice of boundary conditions.

Second, it remains completely open, what kind of excitation we catch
by considering a vortex state in normal mode space.

Finally, it would be interesting to see the changes in the QB
properties, when the norm conservation is lifted. In general we
expect, that this will lead to the generation of higher harmonics,
and new types of resonances. In fact it is exactly resonances due
to higher harmonics, which give the leading order contribution to
the FPU problem. However, for sufficiently narrow optical bands,
higher harmonics will be located outside the optical band. These
resonances will be therefore important, when the optical band is
wide, or when one considers a complex and broader band structure.

{\bf Acknowledgements}
\\
Part of this work was completed while SF was visiting the
Institute for Mathematical Sciences, National University of
Singapore in 2007. KM, OK and MI acknowledge support from RFBR,
grants No.~06-02-16499,~07-02-01404.


\begin{thebibliography}{99}
\bibitem{SPO} $q$-breathers with seed linear modes $q_{0i}=(N+1)/3$,$(N+1)/2$,
$2(N+1)/3$ ($i=\overline{1,d}$) are compact in $q$-space because
these seed modes do not excite the other modes due to the
selection rules (\ref{eq5}).
\end{thebibliography}
\end{document}